\documentclass[twocolumn,showpacs,prb]{revtex4}
\usepackage{bm}
\usepackage{graphicx}
\usepackage{amsmath}
\usepackage{eufrak}
\usepackage{color}
\newcommand{\nix}[1]{}
\begin{document}

\title{Lateral optical anisotropy of type-II interfaces in the
tight-binding approach}
\author{E.L.~Ivchenko}
\author{M.O.~Nestoklon} \email{nestoklon@coherent.ioffe.ru}
\affiliation{A.F.~Ioffe Physico-Technical Institute, Russian
Academy of sciences, St. Petersburg 194021, Russia}
\begin{abstract}
We have developed the tight-binding theory to study electronic and
optical properties of type-II heterostructures CA/C$'$A$'$ grown
from the zinc-blende semiconductors CA and C$'$A$'$ along the
crystallographic direction [001]. The $sp^3s^*$ nearest-neighbor
tight-binding model with allowance for the spin-orbit interaction
is used to calculate the energy states and the in-plane linear
polarization of the spatially-indirect band-edge photoluminescence
of InAs/AlSb and ZnSe/BeTe multi-layered structures. The interface
parameters for a pair of the nonstandard planes C-A$'$ or C$'$-A
are considered as fitting variables. A wide range of these
parameters are shown to allow Tamm-like hole states localized at
the interfaces. The theory leads to giant values of the light
polarization in the both type-II heterosystems in agreement with
existing experimental findings.
\end{abstract}
\pacs{73.21.Fg, 78.67.De}
\maketitle

\section{Introduction}
Experiments on absorption-edge photoluminescence of
InAs/AlSb,\cite{fuchs,berlin}
ZnSe/BeTe\cite{Platonov_1999,Platonov_2000,Platonov_2002},
CdS/ZnSe\cite{gruen} and CdSe/BeTe\cite{Toropov_ssc} multi-layered
structures have revealed a giant in-plane optical anisotropy in
type-II zinc-blende-lattice nanostructures grown along the [001]
crystallographic direction, namely, nonequivalence between the
axes $x
\parallel [1\bar{1}0]$ and $y \parallel [110]$. In our previous
theoretical work\cite{JETP} we have shown that the dominating role
in the nonequivalence of the in-plane axes is played by the
interface bond alignment. In a zinc-blende-lattice bulk
semiconductor, say InAs as a representative of III-V compounds or
ZnSe as a representative of II-VI compounds, any lattice site is a
center of the tetrahedral point symmetry. This means in particular
that if the [001] axis is chosen as a horizontal line then, for
any anion atom, the right-hand-side bonds always lie in the {\it
same} $\langle 110 \rangle$-like plane, say the (110) plane,
whereas the bonds on the left lie in the perpendicular plane
(1$\bar{1}$0). An interface between two materials CA and C$'$A$'$
lacking common anions and cations, A $\neq$ A$^{\prime}$, C $\neq$
C$^{\prime}$, consists of two nonstandard planes containing anions
of one material and cations of another material and forming an
array C-A$'$ either C$'$-A of definitely-oriented nonstandard
chemical bonds. As a result, the relative contributions of the
$p_{x}$- and $p_{y}$-orbitals to the valence-band function near
the interface differ substantially.\cite{JETP} It is this
difference that leads to the observed remarkable anisotropy for
the band-edge, spatially-indirect, radiative processes at the
type-II interface in the polarizations ${\bm e} \parallel x$ and
${\bm e} \parallel y$. The other interface situated vis-a-vis also
induces the in-plane anisotropy. If its chemical bonds are
different then the point symmetry of the structure is $C_{2v}$ and
the anisotropy is retained. The symmetry of an ideal
heterostructure with chemically-identical interfaces is higher
than that of a single heterojunction because it contains a mirror
rotation about the [001] axis by 90$^{\circ}$. In this case the
role of the axes [110] and [1$\bar{1}$0] is interchanged for the
left- and right-hand-side interfaces, their contributions to the
anisotropy cancel each other and these axes become equivalent. A
built-in or external electric field ${\bm E}
\parallel [001]$ breaks the balance and induces the
anisotropy.\cite{Platonov_1999,Platonov_2002} Thus, the polarized
spectroscopy may be suggested as an efficient method to get
information concerning the chemical bonds, intermixing or
reconstruction at the interfaces and the comparative properties of
normal and inverted interfaces.

The calculation of the matrix element of optical transition at a
type-II interface requires the microscopical knowledge of the
electron and hole wave functions near the interface. This
information can be obtained in a microscopic consideration within
the pseudopotential or tight-binding theories rather than by using
the effective-mass approximation or the Kane model. The $sp^3$
tight-binding theory developed in Ref. \onlinecite{JETP} is the
first attempt to describe the giant lateral anisotropy of type-II
interfaces, particularly, of ZnSe/BeTe interfaces. The further
efforts are needed to extend the theory including into
consideration the spin-orbit splitting of the $\Gamma_{15}$ bands
into $\Gamma_8$ and $\Gamma_7$ subbands\cite{Chadi} and additional
atomic orbitals, i.e. $s^*$-orbitals.\cite{Vogl} Moreover, it is
quite well established, at least theoretically, that at the
heterointerface In-Sb in an InAs/AlSb structure there exist
interface, or Tamm-like, states.\cite{Kroemer} In particular, Shen
{\it et al.}\cite{Shen} have performed estimations of the
localization energy for such states in the tight-binding model
neglecting the valence-band spin-orbit splitting in the bulk
compositional materials. Clearly, it is intriguing to analyze the
interplay between the interface and quantum-confined states in
such structures and compare the lateral linear polarization of the
photoluminescence due to both kinds of valence-band states. This
is the main task of the present work, where we have used the
$sp^3s^*$ tight-binding model to calculate the conduction- and
valence-band states in type-II heterostructures taking into
account the spin-orbit interaction, analyze the polarization
properties of band-edge optical matrix elements and obtain the
quantitative results for InAs/AlSb and ZnSe/BeTe heterostructures.
The preliminary results concerning the interface states in
InAs/AlSb heterostructures have been reported in
Ref.~\onlinecite{Nanostr}.

The paper is organized as follows. In Sec.~II we give a brief
description of the $sp^3s^*$ tight-binding model applied to
calculate the conduction- and valence-band states in
nanostructures, discuss the time-inversion symmetry of the model
and go down the list of tight-binding parameters. Section III
contains the derivation of interband matrix elements for the
optical transitions with allowance made for the spin-orbit
interaction. In Sec. IV we present the results of calculations for
InAs/AlSb and ZnSe/BeTe heterosystems including the hole
localization at InAs/AlSb interfaces (Sec.~III A) and the in-plane
linear polarization of the photoluminescence (Secs.~III B and III
C).
\section{Tight-Binding Model Formalism}
First of all we outline a tight-binding theory suitable for the
calculation of interband optical transitions on a type-II
heterojunction. Let us consider a heterostructure grown along the
axis [001]. For the electron states with a zero lateral wave
vector, i.e., for the states with $k_x = k_y = 0$, the electron
wave function in the tight-binding method is written in the form
\begin{equation} \label{56-1}
\psi ({\bm r}) = \sum_{n,\alpha} C^{(\alpha)}_n \phi_{n \alpha}
(x,y,z - z_n)\:.
\end{equation}
Here $\phi_{n \alpha} ({\bm r})$ are the planar atomic orbitals,
$\alpha$ is the orbital state index, $n$ enumerates the atomic
planes, anionic for even $n=2l$ and cationic for odd $n = 2l-1$,
$z_n = na_0/4$ is the position of $n$th atomic plane, and $a_0$ is
the lattice constant of the face-centered cubic lattice. The
planar orbitals are related to the atomic orbitals $\Phi_{n
\alpha}({\bm r})$ by
\[
\phi_{n \alpha} = \sum_{n_1, n_2} \Phi_{n \alpha}({\bm r} - {\bm
a}_n - n_1 {\bm o}_1 -  n_2 {\bm o}_2) \:,
\]
where $n_1, n_2$ are arbitrary integers, ${\bm o}_1 =
(a_0/2)(1,-1,0)$, $ {\bm o}_2 = (a_0/2)(1,1,0)$, and ${\bm a}_n$
is the position of any atom on the $n$-th atomic plane.

We use the $sp^3s^*$ tight-binding model taking into account the
spin-orbit splitting of the $p$-states. Hence, for each value of
$n$, the subscript $\alpha$ in $\phi_{n \alpha}$ runs through ten
values corresponding to ten states $| \Gamma_6, s \rangle $ ($ s =
\pm 1/2) $, $|\Gamma_8, m \rangle$ ($m =3/2, 1/2, -1/2, -3/2$),
$|\Gamma_7, m \rangle$ ($m = \pm 1/2)$, $| \Gamma^*_6, s \rangle $
($ s = \pm 1/2)$. These states can be expressed in terms of $s, p$
and $s^*$ orbitals $S, X, Y, Z, S^*$ as
\begin{eqnarray} \label{basis}
&& \left\vert \Gamma_{6}, 1/2 \right> = \uparrow S\:,\: \left
\vert \Gamma_{6},-1/2 \right> = \downarrow  S \nonumber\:,\\
&&\left\vert \Gamma_{8}, 3/2 \right>= - \uparrow \frac{X + {\rm
i}Y}{\sqrt{2}} \nonumber \:,\\
&&\left\vert \Gamma_{8},1/2 \right> = \sqrt{\frac{2}{3}}\:
\uparrow Z - \downarrow \frac{X+{\rm i}Y}{\sqrt{6}} \nonumber\:,\\
&&\left\vert \Gamma_{8},-1/2 \right> = \sqrt{\frac{2}{3}}\:
\downarrow Z + \uparrow \frac{X-{\rm i}Y}{\sqrt{6}} \nonumber\:,\\
&&\left \vert \Gamma_{8},-3/2 \right> = \downarrow  \frac{X - {\rm
i}Y}{\sqrt{2}} \nonumber \:,\\ &&\left \vert \Gamma_{7},1/2
\right>=\frac{1}{\sqrt{3}} [\uparrow Z+ \downarrow(X+{\rm i}Y) ]\:
,\\ &&\left \vert \Gamma_{7},-1/2 \right>=\frac{1}{\sqrt{3}}\: [
\downarrow Z - \uparrow (X- {\rm i}Y) ] \nonumber\:,\\ &&\left
\vert \Gamma_{6}^*,1/2 \right>=\uparrow S^*\:,\: \left \vert
\Gamma_{6}^*,-1/2 \right>=\downarrow S^* \nonumber\:.
\end{eqnarray}
Hereafter the orbitals $X$ and $Y$ as well as the axes $x, y$ are
oriented along $[1\bar{1}0]$ and $[110]$, respectively.

In the tight-binding method, the wave equation for an electron
with the energy $E$ transforms into a system of linear equations
for the coefficients $C^{(\alpha)}_n$, namely,
\begin{equation} \label{56-2}
\left( E^{\alpha b}_n - E \right) C^{(\alpha)}_n + \frac12
\sum_{n' \neq n , \alpha'} V_{n\alpha,n'\alpha'}
C^{(\alpha')}_{n'} = 0\:.
\end{equation}
Here $E^{\alpha}_n$ are the diagonal atomic energies, and $V_{n
\alpha, n' \alpha'}$ = $V_{n' \alpha', n \alpha}$ are the
off-diagonal matrix elements of tight-binding Hamiltonian for the
pair $n, n'$. In this work we assume the nearest neighbor
approximation, where $V_{n\alpha,n'\alpha'}=0$ for $n\neq n'\pm1$.

Neglecting the spin-orbit interaction, the $sp^3$ nearest-neighbor
tight-binding Hamiltonian in a homogeneous semiconductor crystal
is described by nine parameters $E_{sa}, E_{sc}, E_{pa}, E_{pc}$,
$V_{ss}$, $V_{xx}$, $V_{xy}$, $V_{sa,pc} = V_{pc,sa}$ and
$V_{sc,pa} = V_{pa,sc}$.\cite{JETP} Note that only in this
particular case the notation $x,y$ means the crystallographic axes
[100] and [010]. Following Vogl {\it{et al}}.\cite{Vogl} we
include $s^*$-orbitals adding four other parameters $E_{s^*a}$,
$E_{s^*c}, V_{s^*a,pc}, V_{s^*c,pa}$ and neglecting the transfer
integrals $V_{s^*s^*}, V_{s^*a,sc}, V_{s^*c,sa}$. The spin-orbit
splittings $\Delta_c, \Delta_a$ of the $p$-orbital cation and
anion states complete the list of tight-binding parameters.

Due to the time-inversion symmetry, any electronic state in the
heterostructure with $k_x=k_y=0$ must be doubly degenerate. This
is the so-called Kramers degeneracy. It means that if we have a
solution of the Schr\"odinger equation $\psi$ we can obtain the
second one, $\bar{\psi}$, by applying the time-inversion operator
\begin{equation} \label{time}
\bar{\psi} =\hat{\cal K} \psi \equiv {\rm i} \hat{\sigma}_y
\psi^*\:.
\end{equation}
In the basis (\ref{basis}), the tight-binding expansion
coefficients $C_n^{(\alpha)}, \bar{C}_n^{(\alpha)}$ for the
functions $\psi$ and $\bar{\psi}$ are interrelated by
\begin{equation}
\bar{C}_n^{(\alpha)}=\epsilon_{\alpha} C_n^{(\bar{\alpha})*},
\end{equation}
where the orbitals $\bar{\alpha}$ and $\alpha$ are related by
(\ref{time}), $\epsilon_{\alpha} = 1$ for the orbitals $\alpha$
with the angular momentum components $s,m = 1/2,-3/2$ and
$\epsilon_{\alpha} = -1$ for $\alpha$ with $s,m = - 1/2, 3/2$.

The time-inversion symmetry allows us to separate the states with
$k_x=k_y=0$ into two sets, $A$ and $B$, one with $C_n^{(\alpha)}
=0$ for $s,m = - 1/2, 3/2$ and the other with $C_n^{(\alpha)} =0$
for $s,m = 1/2, - 3/2$. Below we will present explicit equations
for $C^{(\alpha)}_n$ for the first set because the states
corresponding to the set $B$ are easily obtained by using
Eq.~(\ref{time}). Thus, we assume that the wave functions
(\ref{56-1}) are expanded in the basis
\begin{equation}\label{basis_5}
\vert\Gamma_6,1/2\rangle\:\:\:\vert\Gamma_8,-3/2\rangle\:\:,\:\:
\vert\Gamma_8,1/2\rangle\:\:,\:\:\vert\Gamma_7,1/2\rangle\:\:,\:\:
\vert\Gamma_6^*,1/2\rangle\:.
\end{equation}
Presenting the corresponding five coefficients $C_n^{(\alpha)}$ as
a five-component column $\hat{C}_n$ we can write the Schr\"odinger
equation in a matrix form as
\begin{gather}
\hat{V}_{2l-1,2l-2} \hat{C}_{2l-2}^a + \left( \hat{E}_{2l-1} - E
\right) \nonumber \hat{C}^c_{2l-1} + \hat{V}_{2l-1,2l}
\hat{C}_{2l}^a = 0\:, \\ \hat{V}_{2l,2l-1} \hat{C}_{2l-1}^c +
\left( \hat{E}_{2l} - E \right) \hat{C}^a_{2l} + \hat{V}_{2l,2l +
1} \hat{C}_{2l + 1}^c = 0\:. \label{main}
\end{gather}
For clarity, the symbols $\hat{C}_n$ are supplied with an
additional superscript $a$ for anions (even $n$) and $c$ for
cations (odd $n$). Other notations used are as follows:
$\hat{E}_n$ is a diagonal $5\times 5$ matrix with the components
$E_{sc}$, $E_{pc} + \Delta_c/3$, $E_{pc} + \Delta_c/3$, $E_{pc} -
2\Delta_c/3$, $E_{s^*c}$ if $n=2l-1$ and $E_{sa}$, $E_{pa} +
\Delta_a/3$, $E_{pa} + \Delta_a/3$, $E_{pa} - 2\Delta_a/3$,
$E_{s^*a}$ if $n=2l$; $\{ \hat{V}_{n, n'} \}_{\alpha,\alpha'}$ are
$5 \times 5$ matrices of tight-binding cation-anion transfer
integrals $V_{n\alpha,n'\alpha'}$. In particular, the matrix
$\hat{V}_{2l-1,2l}$ is given by
\begin{widetext}
\begin{equation} \label{2l-12l}
\hat{V}_{2l-1,2l} = \frac12 \left[ \begin{array}{ccccc} V_{ss} & 0
& \eta V_{sc,pa}& \xi V_{sc,pa}&0 \\ 0 &  V_{xx} & - \xi V_{xy} &
\eta V_{xy}&0 \\ - \eta V_{sa,pc} &- \xi V_{xy} & V_{xx} & 0 & -
\eta V_{s^*a,pc}\\ - \xi V_{sa,pc} & \eta V_{xy} & 0 & V_{xx} & -
\xi V_{s^*a,pc} \\ 0 & 0 & \eta V_{s^*c,pa}& \xi V_{s^*c,pa}&0
\end{array} \right],
\end{equation}
\end{widetext}
where $\eta = \sqrt{2/3}$, $\xi = \sqrt{1/3}$. The remaining
matrices can be obtained from (\ref{2l-12l}) taking into account
the symmetry and hermicity of the tight-binding Hamiltonian
\begin{eqnarray}
&&\{\hat{V}_{2l-1,2l-2}\}_{\alpha,\alpha'} = (1-2\delta_{\alpha,
\alpha'}) \{ \hat{V}_{2l-1,2l} \}_{\alpha,\alpha'}\:,\nonumber \\
&&\{\hat{V}_{2l,2l-1}\}_{\alpha,\alpha'} = \{ \hat{V}_{2l-1,2l}
\}_{\alpha', \alpha}\:,\\ &&\{ \hat{V}_{2l,2l+1}
\}_{\alpha,\alpha'} = (1-2\delta_{\alpha,\alpha'}) \{
\hat{V}_{2l-1,2l} \}_{\alpha',\alpha}\:.\nonumber
\end{eqnarray}
\subsection{Energy dispersion in a bulk homogeneous semiconductor}
In a three-dimensional semiconductor crystal, the tight-binding
coefficients representing the Bloch solutions with the electron
wave vector ${\bm k}
\parallel [001]$ are written as
\begin{equation} \label{C_exp}
\hat{C}^a_{n}=\hat{C}_a e^{{\rm i} n \phi}\:,\:
\hat{C}^c_{n}=\hat{C}_c e^{{\rm i}n \phi}\:.
\end{equation}
Here $\phi = k a_0 /4$, $\hat{C}_a$ and $\hat{C}_c$ are
independent of $n$ and satisfy the matrix equations
\begin{equation} \label{dispersion}
(\hat{E}_a - E ) \hat{C}_a + \hat{U}_{a} \hat{C}_c = 0\:,\:
\hat{U}_{c} \hat{C}_a + (\hat{E}_c - E)\hat{C}_c = 0 \:,
\end{equation}
\[
\hat{U}_{a} = \hat{U}_{c}^{\dag} = e^{-{\rm i} \phi}
\hat{V}_{0,-1} + e^{{\rm i} \phi} \hat{V}_{0,1}\:.
\]
We took into account that in a periodic system the matrices
$\hat{E}_{2l}, \hat{E}_{2l+1},\hat{V}_{2l, 2l \pm 1}$ etc. are
independent of $l$. Thus, the electron dispersion is determined
from a secular $10 \times 10$ equation.
\subsection{Application to nanostructures}
In the following we consider a type-II heterostructure grown along
the axis [001] and consisting of alternating layers of binary
compounds CA and C$'$A$'$ with different cations and anions. We
assume the conduction-band bottom to be lower in the CA material
and the valence-band top higher in the C$'$A$'$ material (Fig.~1).
The layers are thick enough in order to neglect the formation of
superlattice minibands. Thus, the conduction-band states $\psi_c
({\bm r})$ are approximately calculated for a single CA layer
sandwiched between semiinfinite C$'$A$'$ layers. In the actual
computation procedure the C$'$A$'$-layer thicknesses are taken
finite but large enough to have $\psi_c$ independent on their size
and the external boundary conditions. Similarly, the valence-band
states $\psi_v({\bm r})$ are found for the double-interface
structure CA/C$'$A$'$/CA with thick enough CA layers. By choosing
an appropriate set of the tight-binding parameters for all atoms
in an InAs/AlAs or ZnSe/BeTe structure, see the details below, and
using the standard linear algebra package\cite{cLapack} we can
solve Eqs.~(\ref{main}) and obtain both conduction- and
valence-band tight-binding states. Then we calculate the optical
matrix element due to the indirect photoexcitation $\psi_v \to
\psi_c$ or radiative recombination $\psi_c \to \psi_v$ at the
CA/C$'$A$'$ interface of the periodic heterostructure with
$\psi_c$ being mostly confined within the CA layer and $\psi_v$
confined within the neighboring C$'$A$'$ layer, as shown by
bell-shaped curves in Fig.~1.
\begin{figure}[t]
  \centering
    \includegraphics[width=.4\textwidth]{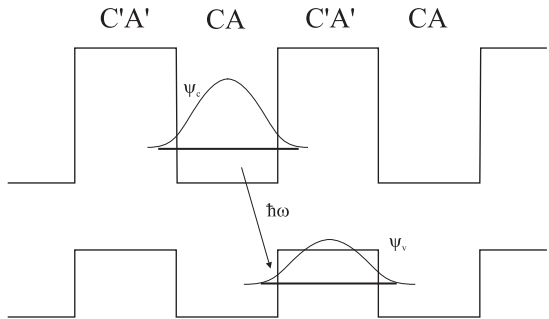}
  \caption{Band diagram of a type-II heterostructure consisting of
  the alternating CA and C$'$A$'$ layers. The higher and lower
  broken lines show the coordinate dependence
of the conduction-band bottom and the valence-band top in the
direction of the growth direction $z$. The arrow
  illustrates the spatially-indirect optical transition.
  }\label{f_bandstructure}
\end{figure}

\subsection{Choice of tight-binding parameters}
The chosen sets of tight-binding parameters for bulk InAs, AlSb,
ZnSe and BeTe are presented in Table \ref{t_params}. For InAs and
AlSb, the parameters are taken from the $sp^3s^*$ model of Klimeck
{\it{et al}}.\cite{Klimeck} For bulk ZnSe, the 15 tight-binding
parameters of the $sp^3s^*$ model can be borrowed from
Refs.~\onlinecite{Bertho91}, \onlinecite{Bertho93} and
\onlinecite{Dierks}. In the main calculation we use the parameters
from Ref.~\onlinecite{Bertho93}.
The tight-binding parameters for BeTe are the result of our fit of
the bandstructure data presented in Table \ref{t_fit}. The band
energies at the $\Gamma$ and $X$ points are calculated by using
Eqs.~(\ref{dispersion}), and the effective masses are evaluated by
using equations for $m_{e}, m_{hh}, m_{lh}, m_{so}$ derived in
Ref.~\onlinecite{Boykin}. The parameters evidently differ from
those used in Refs.~\onlinecite{JETP} and \onlinecite{Zhu}, where
the spin-orbit interaction is neglected.

In Table \ref{t_params} the diagonal energies refer to the
valence-band top, $E_{v, \Gamma_8}$, of each bulk material. The
diagonal energies $E_{\alpha b}$ used in the calculation for AlSb
and BeTe differ from those in the table by the valence band offset
$V$ referred to InAs and ZnSe, respectively. In accordance with
estimations in Refs.~\onlinecite{Nakagawa,Dandrea,Shen,Wei} the
offset $V \equiv \Delta E_v (\mbox{InAs} \to \mbox{AlSb}) =$ 0.10
eV either 0.15 eV is taken for the InAs/AlSb heteropair. For the
offset $\Delta E_v (\mbox{BeTe} \to \mbox{ZnSe})$ we use the value
0.95 eV.\cite{Platonov_1999}

As far as the interface parameters are concerned, for a pair of
interface atomic planes C-A$'$ or C$'$-A we start from the average
values $E_{sa}(\mbox{CA$'$}) = [ E_{sa}(\mbox{A}) + E_{sa}
(\mbox{A}')]/2$, $V_{xx}(\mbox{CA}') = [ V_{xx}(\mbox{CA}) +
V_{xx}(\mbox{C}'\mbox{A}')]/2$, etc. Then the sets $E_{\alpha
b}(\mbox{CA}')$ and $E_{\alpha b}(\mbox{C}'\mbox{A})$ are shifted
by constant energies in order to satisfy the conditions
\begin{eqnarray}
E_{v \Gamma_8}(\mbox{CA}') - E_{v \Gamma_8}(\mbox{CA}) = V'(\mbox{CA}')\:,
\nonumber\\
E_{v \Gamma_8}(\mbox{C}'\mbox{A}) - E_{v \Gamma_8}(\mbox{CA}) =
V'(\mbox{C}'\mbox{A})\:, \nonumber
\end{eqnarray}
where $V'(\mbox{C}\mbox{A}')$ and $V'(\mbox{C}'\mbox{A})$ are
variable parameters of the theory. Shen {\it et al.}\cite{Shen}
used the value $V'$(InSb)$=0.75$ eV while the first-principles
all-electron band structure method gives $V'$(InSb)$=0.50$ eV.
\cite{Wei} In this paper we do not confine ourselves to a
particular value of $V'$(InSb) and assume it to lie in the range
$0.1\div 0.75$ eV. As for the offset $V'$(ZnTe), we used in our
calculation three different values $0.5$, $0.75$ and $1.0$ eV
bearing in mind that Wei and Zunger~\cite{Wei} obtained
$V'$(ZnTe)=0.73 eV. In fact, the bulk semiconductors InAs and AlSb
(or ZnSe and BeTe) have very close lattice constants which differ
remarkably from the lattice constant of bulk InSb (or ZnTe). Due
to this lattice mismatch, the InSb and ZnTe interface atomic
planes are strongly strained and the related parameters should be
quite different from those of unstrained bulk InSb and ZnTe. We
take this uncertainty into account by considering the interface
off-diagonal matrix element $V_{xy}$ as an additional variable
parameter.

\begin{table}
\caption{Tight-binding parameters used in the calculations. The
diagonal energies are referred to the $\Gamma_8$ valence-band top
of the corresponding bulk material.}\label{t_params}
\begin{ruledtabular}
\begin{tabular}{ccccc} 
           &   InAs       &      AlSb    &  ZnSe &  BeTe \\ \hline
$E_{sa}$&     -9.57566    &   -4.55720   & -10.19& -9.907\\
$E_{pa}$&      0.02402    &    0.01635   &  0.06 &  0.58 \\
$E_{sc}$&     -2.21525    &   -4.11800   &  0.76 &  2.04 \\
$E_{pc}$&      4.64241    &    4.87411   &  7.22 &  3.96 \\
$E_{s^*a}$&    7.44461    &    9.84286   &  10.0 &  8.0  \\
$E_{s^*c}$&    4.12648    &    7.43245   &  12.0 &  9.06 \\ \hline
$V_{ss}$&     -5.06858    &    -6.63365  & -5.17 & -6.00 \\
$V_{xx}$&      0.84908    &    1.10706   &  1.22 &  1.96 \\
$V_{xy}$&      4.68538    &    4.89960   &  5.48 &  5.5  \\
$V_{sa,pc}$&   2.51793    &    4.58724   &  5.41 & -1.0  \\
$V_{sc,pa}$&   6.18038    &    8.53398   &  6.62 &  7.5  \\
$V_{s^*apc}$&  3.79662    &    7.38446   &  5.63 &  2.8  \\
$V_{s^*cpa}$&  2.45537    &    6.29608    &  5.75 &  5.5  \\
\hline
$\Delta_a$&    0.38159    &    0.70373  &  0.43 &  1.1  \\
$\Delta_c$&    0.37518    &    0.03062  &  0.038&  0.26  \\
\end{tabular}
\end{ruledtabular}
\end{table}

\begin{table}[b]
\caption{The band edges referred to the $\Gamma_8$ valence-band
top and effective masses of BeTe given in literature (target) and
calculated by using the tight-binding (TB) parameters of Table
\ref{t_params}. }\label{t_fit}
\begin{ruledtabular}
\begin{tabular}{ccc||ccc} 
                &   BeTe & BeTe & & BeTe & BeTe\\
&TB&target&&TB&target\footnotemark[1]\\\hline
$\Gamma_6^c$    & 4.53 &    4.53\footnote{Ref.~\cite{Nagelstrasser}}   &$m_e$     &-0.04 & $<$ 0    \\
$\Gamma_7^v$    &-0.96 &   - 0.96\footnotemark[1]  &$m_{hh}$  & 0.34 &  0.34    \\
$\Gamma_7^c$    & 4.6  &     4.64\footnote{Ref.~\cite{Fleszar}}  &$m_{lh}$  & 0.24 &  0.23 \\
$\Gamma_8^c$    & 5.0  &   4.99\footnotemark[2]    &$m_{so}$  & 0.33 &  0.40  \\
$X^c_6$         & 2.6  &    2.7\footnotemark[1]    &&&
\end{tabular}
\end{ruledtabular}
\end{table}

\section{Interband matrix element of the optical
transition}\label{sect_Interband_martel} The matrix element of the
optical transition for the photon polarization ${\bm e}$ is
proportional to the matrix element of the scalar product of the
velocity operator and ${\bm e}$. In order to express the matrix
elements in terms of the expansion of $\psi$ in planar orbitals,
we must first express the velocity operator in the basis of atomic
orbitals. The atomic sites are completely determined by the
position of the atom ${\bm R} = {\bm a} + {\bm \tau}_b$ specified
by the location of an elementary cell, ${\bm a}$, and the location
${\bm \tau}_b$ of the $b$ atom within the cell. Then the
tight-binding Hamiltonian is determined by the matrix elements
$H_{\alpha' \alpha} (\bm{R}', \bm{R})$.

The expression for matrix elements of the velocity operator can be
found by using the relation
\begin{equation} \label{velocity}
\hat{\bm{v}} = \frac{\rm i}{\hbar} (H \bm{r}- \bm{r} H)
\end{equation}
between the velocity and coordinate operators, taking the
Hamiltonian $H$ in the form $H_{\alpha' \alpha} (\bm{R}', \bm{R})$
and introducing the matrix elements of the coordinate operator
$\bm{r}_{\alpha' \alpha}(\bm{R}',\bm{R})$. As a rule, only
intrasite matrix elements
\begin{equation}
{\bm r}_{\alpha' \alpha} (\bm{R}', \bm{R}) = (\bm{R} \:
\delta_{\alpha' \alpha} + \bm{r}_{\alpha' \alpha})\: \delta_{{\bm
{R}}', {\bm{R}}}
\end{equation}
are taken into account, see Refs.~\onlinecite{cruz} and
\onlinecite{pedersen} and the bibliography therein, where the
contribution
\[
{\bm r}_{\alpha' \alpha} = \langle \bm{R}, \alpha' | \bm{r} -
\bm{R} | \bm{R}, \alpha \rangle
\]
describes inter-orbital transitions within a single atomic site.
Here we use the theory developed in Ref.~\onlinecite{ram} where
${\bf r}_{\alpha' \alpha}$ are assumed to vanish and the optical
transitions are uniquely determined by the tight-binding
parameters. Then, one obtains for the velocity operator
\[
{\bf v}_{\alpha' \alpha}({\bm R}', {\bm R}) = \frac{\rm
i}{\hbar}\: (\bm{R} - \bm{R}')\:H_{\alpha' \alpha}({\bm R}', {\bm
R})\:.
\]
It is seen that, according to this theory, the intra-atomic terms
are equal to zero, and the inter-atomic terms are directed along
the vector ${\bm R} - {\bm R}'$, i.e., along the chemical bond
between the atoms $\bm{R}$ and $\bm{R}'$. In this case, the
inter-atomic transitions between the planes $2l, 2l - 1$ and $2l,
2l + 1$ cause the emission of photons polarized in the $x$ and $y$
directions, respectively.

In the $sp^3$ tight-binding approach without spin-orbit
interaction an expression for interband matrix element of the
optical transition is \cite{JETP}
\begin{widetext}
\begin{equation} \label{matrel}
M_j = {\rm i} \frac{a_0}{4 \hbar} \sum_l V^j_l \:,
\end{equation}
\begin{eqnarray}
V^{x}_l &=& V_{sa,pc} C^{sa*}_{2l} C^{p_{x} c}_{2l-1} + V_{pa,sc}
C^{sc*}_{2l-1} C^{p_{x}a}_{2l} - V_{xy} \left( C^{p_z a *}_{2l}
C^{p_{x} c}_{2l-1} - C^{p_{z} c *}_{2l-1} C^{p_{x}
a}_{2l} \right) \:, \nonumber\\
V^{y}_l &=&   V_{sa,pc} C^{sa*}_{2l} C^{p_{y} c}_{2l+1} +
V_{pa,sc} C^{sc*}_{2l + 1} C^{p_{y}a}_{2l} + V_{xy} \left( C^{p_z
a *}_{2l} C^{p_{y} c}_{2l+1} - C^{p_{z} c *}_{2l+1} C^{p_{y}
a}_{2l} \right)\:. \nonumber
\end{eqnarray}
\end{widetext}
Here $M_j$ is the interband matrix element of the velocity
operator $\hat{v}_j$ ($j = x, y$), $V^{x}_l$ is the contribution
to $M_x$ from inter-atomic transitions between the anion plane
$2l$ and the cation plane $2l-1$, $V^y_l$ is a similar
contribution to $M_y$ from the $2l$ and $2l+1$ planes, $C^{s b}_n$
and $C^{p_z b}_n$ are the coefficients describing the admixture of
$s$- and $p_z$-orbitals in the expansion (\ref{56-1}) for an
electron state in the lowest conduction band $\Gamma_1$, $C^{p_j
b}_n$ are the $p_j$-orbital coefficients for the hole states in
the valence band, $V_{sa,pc}, V_{pa,sc}$ and $V_{xy}$ are the
above anion-cation transfer integrals. It should be mentioned that
Eq.~(\ref{matrel}) describes the photon absorption. The emission
matrix elements are obtained by the complex conjugation of $M_j$.

The spin-orbit interaction included, one has to consider four
interband optical transitions $v,A \to c,A$, $v,B \to c,B$, $v,A
\to c,B$ and $v, B \to c, A$ for each pair of the conduction ($c$)
and valence ($v$) subbands, where $A,B$ denote the
Kramers-conjugate sets of states. In the polarization ${\bm e}
\perp z$ the first two transitions are forbidden while the
absolute values of the matrix elements for the two latter
transitions coincide. Hence, it suffices to present an equation
only for the $v, B \to c, A$ matrix element. For ${\bm e}
\parallel x$ and ${\bm e} \parallel y$, it reads, respectively,
\begin{equation} \label{Vx}
V^{x}_l= \hat{C}^{c,A \dag}_{2l-1} \hat{V}_{x1} \hat{C}^{v,
B}_{2l} +  \hat{C}^{c,A \dag}_{2l} \hat{V}_{x2} \hat{C}^{v,
B}_{2l-1}\:,
\end{equation}
\[
V^{y}_l= \hat{C}^{e,A \dag}_{2l+1} \hat{V}_{y1} \hat{C}^{v,
B}_{2l} + \hat{C}^{e, A \dag}_{2l} \hat{V}_{y2} \hat{C}^{v,
B}_{2l+1} \:.
\]
Five coefficients $C^{(\alpha)}_n$ for the initial valence and
final conduction states are represented here by five-component
columns $\hat{C}^{v,B}_n$ and $\hat{C}^{c,A}_n$. The matrices
$\hat{V}_{x1}\:,\:\hat{V}_{y1}$ are given by
\begin{widetext}
\begin{equation} \label{Vx_1}
\hat{V}_{x1}=\frac{1}{\sqrt{2}} \left[\begin{array}{ccccc}
0 & -V_{sc,pa} & \xi V_{sc,pa} & -\eta V_{sc,pa} & 0 \\
-V_{sa,pc} & 0 & \eta V_{xy} & \xi V_{xy} & -V_{s^*a,pc}\\
\xi V_{sa,pc} & -\eta V_{xy} & 0 & -V_{xy} & \xi V_{s^*a,pc}\\
-\eta V_{sa,pc} & -\xi V_{xy} & V_{xy} &0 & -\eta V_{s^*a,pc}\\
0 & -V_{s^*c,pa} & \xi V_{s^*c,pa} & -\eta V_{s^*c,pa} & 0 \\
\end{array}\right]\:,
\end{equation}
\begin{equation}\label{Vy_1}
\hat{V}_{y1}=\frac{\rm i}{\sqrt{2}} \left[\begin{array}{ccccc}
0 & -V_{sc,pa} & -\xi V_{sc,pa} & \eta V_{sc,pa} & 0 \\
-V_{sa,pc} & 0 & -\eta V_{xy} & -\xi V_{xy} & -V_{s^*a,pc}\\
-\xi V_{sa,pc} & \eta V_{xy} & 0 & -V_{xy} & -\xi V_{s^*a,pc}\\
\eta V_{sa,pc} & \xi V_{xy} & V_{xy} &0 & \eta V_{s^*a,pc}\\
0 & -V_{s^*c,pa} & -\xi V_{s^*c,pa} & \eta V_{s^*c,pa} & 0 \\
\end{array}\right]\:,
\end{equation}
\end{widetext}
while the two other matrices are obtained by the transposition,
namely, $\hat{V}_{x2} = \tilde{\hat{V}}_{x1}$, $\hat{V}_{y2} =
\tilde{\hat{V}}_{y1}$.

In the polarization ${\bm e}\parallel z$, for the chosen sets of
states the inter-set optical transitions $A \leftrightarrow B$ are
forbidden while the transitions $v,A \to c,A$ and $v,B \to c,B$
have equal probability rates. For the sake of brevity we omit an
expression for $M_z$ similar to Eqs.~(\ref{matrel}, \ref{Vx}).
\section{Results and Discussion}
Conduction- and valence-band states have been calculated with the
aid of an original computer program using cLapack\cite{cLapack}
which allows to solve the equation set (\ref{main}). Interface
states are identified as those with energies lying above the
valence band top of the C$'$A$'$ material and below the conduction
band bottom of the CA material, see Fig.~1. As mentioned above,
each calculation was performed for a CA (or C$'$A$'$) layer of
finite thickness $L$ sandwiched between thick layers C$'$A$'$ (or
CA) so that both interface and quantum-confined states could be
calculated simultaneously. The localization energy at a single
heterointerface could be found as a nonzero limit of the
localization energy with increasing the width $L$.  In contrast,
the energy of quantum-confined states tends to zero as $ L^{-2}$.
\subsection{Hole localization at InAs/AlSb interfaces}
Figures 2 and 3 present our tight-binding calculations of the two
lowest hole states in a three-layer structure InAs/AlSb/InAs with
the InSb-like interfaces and a 60-\AA--thick AlSb layer (40
monoatomic layers, or 20 monomolecular layers). For the InAs/AlSb
valence band offset $V$, we took 0.1 eV. We define the
localization energy $\varepsilon$ of a hole state as the
difference between the energy $E$ (in the electron representation)
and the AlSb valence-band top, as shown in the inset in Fig.~2.
Therefore, values of $\varepsilon$ are positive for hole interface
states and negative for states quantum-confined within the whole
AlSb layer. One can see from Figs.~2, 3 that there exists a wide
range of interface parameters allowing interface, or Tamm-like,
hole states. The analysis shows that $V_{xy}$ and $V'$ are those
two parameters which have the strongest influence on
$\varepsilon$. Variation of the diagonal energies $E_{\alpha b}$
and the transfer integrals $V_{ss}$, $V_{pp}$ also changes the
level position. However, these parameters affect the localization
energy mostly through a change of the InSb valence-band offset
induced by their variation. This explains why we show in Figs.~2
and 3 the dependence of $\varepsilon$ upon $V'$ and $V_{xy}$
keeping other parameters constant.

\begin{figure}[t]
  \centering
    \includegraphics[width=.45\textwidth]{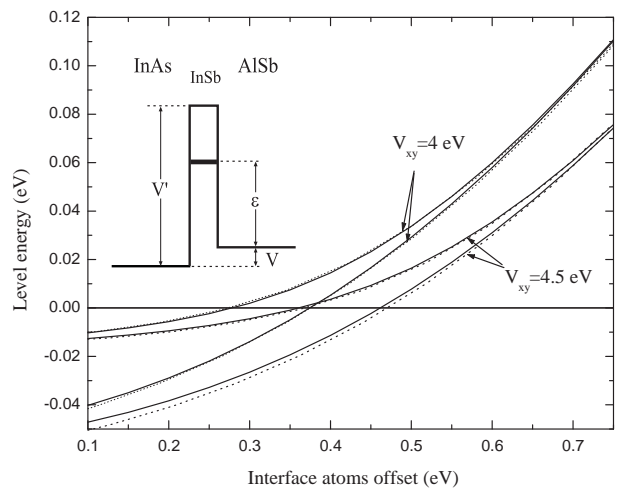}
\caption{Localization energy for symmetrical and antisymmetrical
interface hole states in an InAs/AlSb multilayered structure with
the In-Sb interfacial bonds vs. InSb valence-band offset, $V'$,
for two different values of the interface off-diagonal
tight-binding parameter $V_{xy}= 4.0$ eV and 4.5 eV (solid
curves). Dotted curves represent analytic results in the
envelope-function approximation with a $\delta$-like interface
potential. The inset shows the effective band diagram of the
InSb-like interface.} \label{f_IS_offset}
\end{figure}

To interpret the results physically, we have applied the
envelope-function approximation and simulated the interface
localization by a single-quantum-well structure of the thickness
$L$ with a band offset $V$ and a pair of $\delta$-functions of
equal strength at the interfaces. Thus, the hole potential energy
$V(z)$ as a function of $z$ is assumed to have the form
\[
V \theta\left( |z| - \frac{L}{2} \right) - U a_0 \left[ \delta
\left( z - \frac{L}{2} \right) + \delta \left( z + \frac{L}{2}
\right) \right]\:,
\]
where $\theta(x)$ is the Heaviside step function, the origin $z=0$
is chosen in the well center, and the lattice constant $a_0$ is
introduced in order to have the energy units for the factor $U$.
In the following we take $a_0=6.08$ \AA. Obviously, interface
states with $\varepsilon > 0$ can exist if $U$ is positive and
exceeds some critical value. Two other model parameters are the
heavy-hole effective masses $m_1$ and $m_2$ in bulk AlSb and InAs,
respectively. Their values are taken from our tight-binding
estimations for the bulk compositional materials. It should be
noted (see, e.g., Ref.~\onlinecite{Volkov}) that, in the
effective-mass approximation, the role of the above $\delta$-like
potential is equivalent to the inclusion of an additional term
proportional to $U$ in the boundary conditions for the hole
envelope function $\varphi(z)$, namely,
\[
\varphi_1 \left( \pm \frac{L}{2} \right) = \varphi_2 \left( \pm
\frac{L}{2} \right) \equiv \varphi\left( \pm \frac{L}{2}
\right)\:,
\]
\[
\frac{1}{m_1}\: \varphi'_1 \left( \pm \frac{L}{2} \right) =
\frac{1}{m_2}\: \varphi'_2 \left( \pm \frac{L}{2} \right) \pm
\frac{2 U a}{\hbar^2}\: \varphi \left( \pm \frac{L}{2} \right)\:.
\]
Here the subscripts 1 and 2 denote for short the materials InAs
and AlSb.

Due to the inversion symmetry of the potential energy $V(z)$, the
functions $\varphi(z)$ have a definite parity. For the interface
states, the even or odd solutions can be written in the form $C
\cosh{(\mbox{\ae} z)}$ or $C \sinh{(\mbox{\ae} z)}$ inside the
well, and
\[
C \cosh{ \left( \frac{\mbox{\ae} L}{2} \right) } \exp{\left[ -
\mbox{\ae}' \left(|z| - \frac{L}{2} \right) \right]}
\]
or
\[
C \:\mbox{sign}\{ z \}\: \sinh{ \left( \frac{\mbox{\ae} L}{2}
\right)} \exp{\left[ - \mbox{\ae}' \left(|z| - \frac{L}{2} \right)
\right]}
\]
in the barriers, where
\begin{equation}
\label{aeae'} \mbox{\ae} = \sqrt{2 m_1 \varepsilon/\hbar^2}\:,\:
\mbox{\ae}' = \sqrt{2 m_2(\varepsilon+V)/\hbar^2}\:,
\end{equation}
and $C$ is the normalization coefficient. The localization energy
$\varepsilon$ satisfies the following transcendental equations
\begin{eqnarray} \label{energylev}
\mbox{\ae}' \left[ 1 + \eta\: \tanh{(\mbox{\ae} L/2)} \right] =
\beta \hspace{4 mm} \mbox{(for even solution)} \\  \mbox{\ae}'
\left[ 1 + \eta\: \coth{(\mbox{\ae}L/2)} \right] = \beta \hspace{4
mm} \mbox{(for odd solution)}\:, \nonumber
\end{eqnarray}
where $\eta = (m_1 \mbox{\ae}/m_2 \mbox{\ae}')$, $\beta = 2 m_2
a_0 U / \hbar^2$.

In the particular case of zero offset and equal effective masses,
Eqs.~(\ref{energylev}) reduce to those for the simple
one-dimensional model of a diatomic molecule.\cite{Animalu} For
very wide wells, $\tanh{(\mbox{\ae} L/2)}$, $\coth{(\mbox{\ae}
L/2)} \to 1$ and Eqs.~(\ref{energylev}) turn into the equation
\[
\mbox{\ae}' \left( 1 + \eta \right) = \beta
\]
for the hole interface states at a single heterointerface. For
small values of $\exp{(- \mbox{\ae} L)}$, the even-odd splitting
can be estimated as
\begin{equation}\label{delta_e}
\delta \varepsilon = \frac{8
\varepsilon_0\mbox{\ae}'_0}{\mbox{\ae}'_0 + \mbox{\ae}_0} \exp{(-
\mbox{\ae}_0 L)}\:,
\end{equation}
where the subscript $0$ refers to the values of $\varepsilon,
\mbox{\ae}, \mbox{\ae}'$ related to a single-interface state.

\begin{figure}[t]
  \centering
    \includegraphics[width=.5\textwidth]{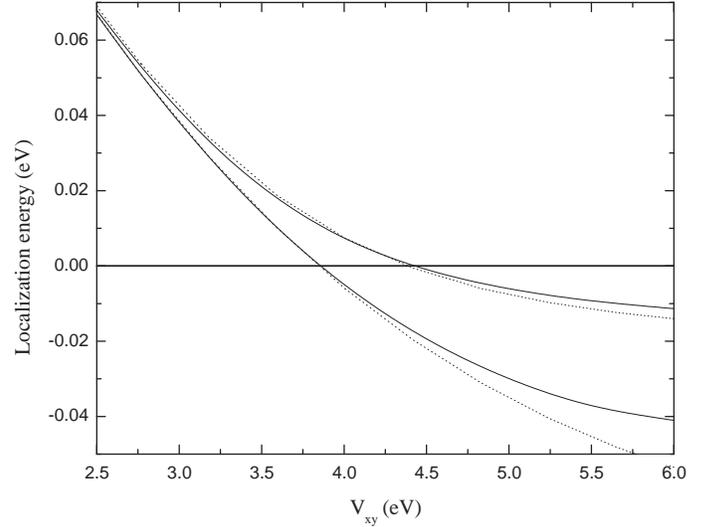}
\caption{Localization energy for the hole interface states as a
function of interfacial tight-binding parameter $V_{xy}$.
Interface-atoms valence-band offset $V'$ is 0.35 eV (solid
curves). Dotted curves are analytic results obtained in the
envelope-function approximation with a $delta$-like interface
potential. }\label{f_IS_Vxy}
\end{figure}

We have tried to fit the tight-binding curves in Figs.~2 and 3 by
choosing a particular form of dependence of $U$ on $V'$ and
$V_{xy}$. For simplicity, we have assumed the linear dependence
\begin{equation} \label{ucc}
U = U_0 + c_1 V' + c_2 V_{xy}\:,
\end{equation}
where $U_0, c_1, c_2$ are fitting parameters. Surprisingly, this
simple assumption has proved to work. The dotted curves in
Figs.~2, 3 represent the results obtained in the envelope-function
approach by using the following set: $U_0 = 26.99$ eV, $c_1 =
33.55, c_2 = - 6.36$. One can see that this approach successfully
simulates the behavior of two lowest hole states in the wide range
of $V'$ and $V_{xy}$. For $V' = 0.35$ eV and $V_{xy} = 4$ eV, the
simplified exponential description of the even-odd splitting by
Eq.~(\ref{delta_e}) is certainly justified in the structures with
the layer thickness $L$ exceeding 60 \AA\mbox{} provided
$\varepsilon_0 \gtrsim 0.015$~eV.

An increase in $V_{xy}$ or a decrease in $V'$ lowers the effective
interface potential and  transforms the pair of interface hole
states into the two lowest quantum-confined hole states with
negative $\varepsilon$. Within a certain two-dimensional area of
$V_{xy}$ and $V'$ the values of $\varepsilon$ for the even and odd
states have opposite signs. This is a transitional area from the
interface localization to quantum confinement. In this case it is
possible that, in a wide well, interface states are absent while,
with decreasing $L$, there appears an even interface-induced state
with $\varepsilon > 0$ which is, in fact, not attached to the
interfaces but rather spread over the whole AlSb layer. In other
words, the existence of a hole state inside the bandgap of the
heterostructure can depend not only on interface properties but
also on the well width $L$.

Since $|c_2| \ll c_1$ we conclude that, as compared with $V_{xy}$,
the offset $V'$ has the much stronger influence on the
localization energy. On the other hand, the parameter $V_{xy}$
plays a more crucial role in the polarization properties of the
vertical band-edge photoluminescence, as will be seen in the next
section. Thus, the above analysis is a clear and unambiguous
indication that the two hole states with $\varepsilon > 0$
represented in Figs.~\ref{f_IS_offset}, \ref{f_IS_Vxy} are
admixtures of the left- and right-hand-side interface states.

Of course, the symmetry analysis can be applied as well to the
tight-binding Hamiltonian and tight-binding solutions. A
(001)-grown InAs/AlSb/InAs structure with the symmetrical
InSb-like interfaces has the $D_{2d}$ point symmetry with the
center $O$ of the point transformations located at any atomic site
in the central cation plane $n_0$. We recall that the $D_{2d}$
group contains a mirror-rotation axis $S_4$. The two lowest hole
states of Figs.~2, 3 have a heavy-hole-like nature with an
admixture of light-hole, spin-orbit-split and conduction-band
states. Therefore, the parity $p= \pm$ of the envelope function
$\varphi(z)$ with respect to the operation $S_4$ coincides with
that of the coefficients $C_n^{(\Gamma_8, - 3/2)}$ or
$C_n^{(\Gamma_8, 3/2)}$ in the expansion (\ref{56-1}) over the $A$
or $B$ basis set, namely,
\begin{equation} \label{parityc}
C_{n - n_0}^{(\Gamma_8, \pm 3/2)} = p\: C_{n_0 - n}^{(\Gamma_8,
\pm 3/2)}\:.
\end{equation}
The coefficients $C_n^{(\Gamma_6, \pm 1/2)}$, $C_n^{(\Gamma^*_6,
\pm 1/2)}$ behave in the same way whereas the parity of
$C_n^{(\Gamma_8, \pm 1/2)}$ and $C_n^{(\Gamma_7, \pm 1/2)}$  is
opposite to (\ref{parityc}). These symmetry considerations are in
a complete agreement with the parities of microscopic solutions
obtained in our tight-binding calculations.

An increase in the InAs/AlSb valence-band offset, $V$, suppresses
the localizing effect of the In-Sb interface. The computation
shows that, at $V = 0.15$ eV, the lowest valence-band states are
pushed out beyond the heterostructure band gap and the energy
$\varepsilon$ takes on negative values within the relevant range
of $V'$ and $V_{xy}$.
\subsection{Optical properties of InAs/AlSb heterostructures}
\begin{figure}[t]
  \centering
    \includegraphics[width=.45\textwidth]{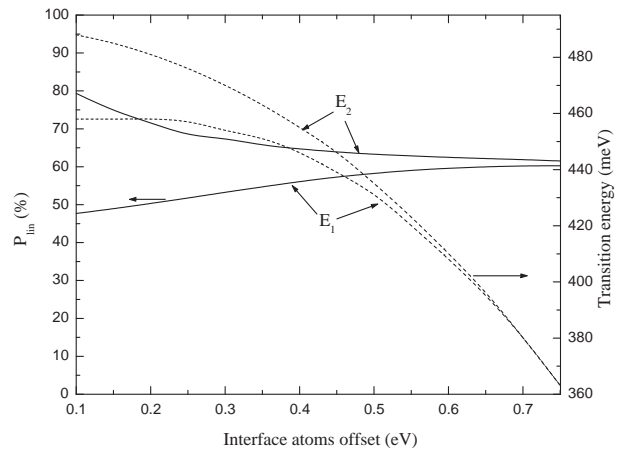}
\caption{Transition energy ({\it dashed}) and linear polarization
({\it solid}) of the indirect radiative recombination at an
60\AA/60\AA\mbox{} InAs/AlSb heterointerface as a function of the
In-Sb interface-atoms valence-band offset $V'$ for the interface
off-diagonal tight-binding parameter $V_{xy}$=4.0
eV.}\label{f_IS_Pol_offset}
\end{figure}
\begin{figure}[t]
  \centering
    \includegraphics[width=.45\textwidth]{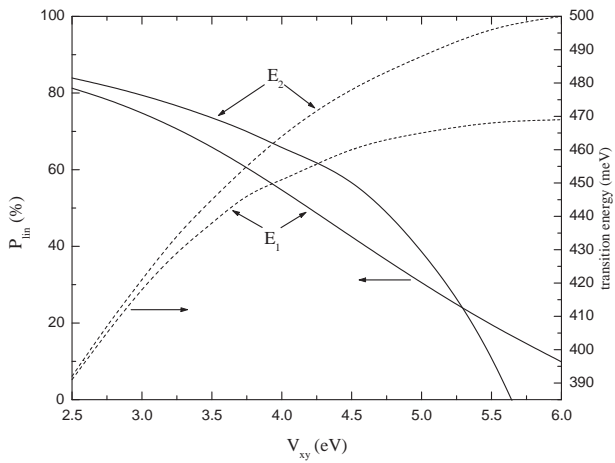}
  \caption{Transition energy ({\it dashed}) and linear polarization
({\it solid}) of indirect radiative-recombination at a
60\AA/60\AA\mbox{} InAs/AlSb heterointerface as a function of the
interface off-diagonal tight-binding parameter $V_{xy}$. The
calculation is performed for for the interface-atoms valence-band
offset $V'$ = 0.35 eV.}\label{f_IS_Pol_Vxy}
\end{figure}

Figures \ref{f_IS_Pol_offset}, \ref{f_IS_Pol_Vxy} show the
transition energy and in-plane linear polarization of the indirect
photoluminescence at a particular interface as functions of the
interface parameters $V'$ and $V_{xy}$. The conduction-band states
are calculated for a three-layer structure AlSb/InAs/AlSb with a
60-\AA-thick inner InAs layer and the InSb-like interfaces. The
transition energies are defined by
\begin{equation} \label{tren}
E_{1,2} = E^{(i)}_g + E_{e1} - \varepsilon_{\pm}\:,
\end{equation}
where the indirect band gap $E^{(i)}_g$ equals the difference,
$E_{c \Gamma_6}(\mbox{InAs}) - E_{v \Gamma_8}(\mbox{AlSb})$,
between the the conduction-band bottom in InAs and the
valence-band top in AlSb, $E_{e1}$ is the quantum-confinement
energy of an electron in the lowest conduction subband $e1$, and
$\varepsilon_{\pm}$ is the hole localization energy for the even
and odd solutions, respectively. The degree of linear polarization
is given by
\begin{equation} \label{polar}
P_{\rm lin} = \frac{|M_x|^2 - |M_y|^2}{|M_x|^2 + |M_y|^2}\:,
\end{equation}
where the matrix elements $M_x, M_y$ are introduced in
Eq.~(\ref{matrel}). For their calculation the envelope-function
approximation is unusable, and one needs a microscopic description
of the $X$- and $Y$-orbital admixture in the electron wave
function (\ref{56-1}).

As compared with $\varepsilon$, the transition energy increases
faster with increasing the parameter $V'$  because the latter also
affects the electron confinement energy $E_{e1}$. On the other
hand, the energy difference $E_2 - E_1$ of course coincides with
$\varepsilon_+ - \varepsilon_-$ as it follows from
Eq.~(\ref{tren}). One can conclude from
Figs.~\ref{f_IS_Pol_offset}, \ref{f_IS_Pol_Vxy} that the linear
polarization is very sensitive to the parameter $V_{xy}$ and
rather stable to a variation of $V'$. This is in line with the
symmetry considerations. The energy $V'$ is a uniaxial invariant
whereas the transfer integral $V_{xy}$ is an interface parameter
governing the in-plane anisotropy expressed in a non-equivalence
between the $[1\bar{1}0]$ and [110] directions. It should be noted
that the same linear polarization is expected for a recombining
exciton formed by an electron and hole confined within the
neighboring CA and C$'$A$'$ layers.

Let us now discuss the light polarization taking into account that
the radiative process can occur at both normal and inverted
interfaces. In an ideal structure of the $D_{2d}$ symmetry, the
photons emitted in the two processes have the same energy, and
their contributions to the photoluminescence intensity coincide.
In this case, as mentioned in the introduction, the linear
polarization vanishes, and the information on the optical
anisotropy of the individual interface is hidden. If by any reason
the photoluminescence intensities $I_i$, $I_n$ related to the
corresponding interface contributions are different, then the
observed degree of linear polarization is given by
\begin{equation}
\bar{P}_{\rm lin}=(2 \zeta - 1) P_{\rm lin}\;,
\end{equation}
where $\zeta = I_n / (I_n + I_i)$, and $P_{\rm lin}$ is the linear
polarization (\ref{polar}) referred to the normal interface.

Experimental studies of spatially direct and indirect
photoluminescence from (001)-grown InAs/AlSb heterostructures are
presented by Fuchs {\it{et al.}}\cite{fuchs,berlin,fuchs1} The
shutter sequence during the MBE growth promoted InSb-like
interfaces which was confirmed via Raman measurements. The
polarization-resolved spatially-indirect photoluminescence of all
samples showed a strong optical in-plane anisotropy. In
particular, this was the case for a multilayered heterostructure
that contained two InAs 75-\AA-thick layers separated by a
50-\AA-thick AlSb layer and surrounded by two other, thick enough,
AlSb layers. For the vertical photoluminescence along the growth
direction, the linear polarization $P_{\rm lin}$ was as high as
$\sim 60$~\%. We assume that, in the experiment \cite{berlin} the
photoluminescence originated from the radiative recombination at
one interface, say, due to a slight asymmetry of the
heterostructure and a preferential occupation by photoholes of the
states at a particular interface. Then according to
Figs.~\ref{f_IS_Pol_offset}, \ref{f_IS_Pol_Vxy} our theory
explains the measured high values of $\bar{P}_{\rm lin}$ provided
$V_{xy} < 4$ eV for $V' = 0.35$ eV or $V' > 0.5$ for $V_{xy} = 4$
eV.

\begin{figure}[t]
\centering
\includegraphics[width=.45\textwidth]{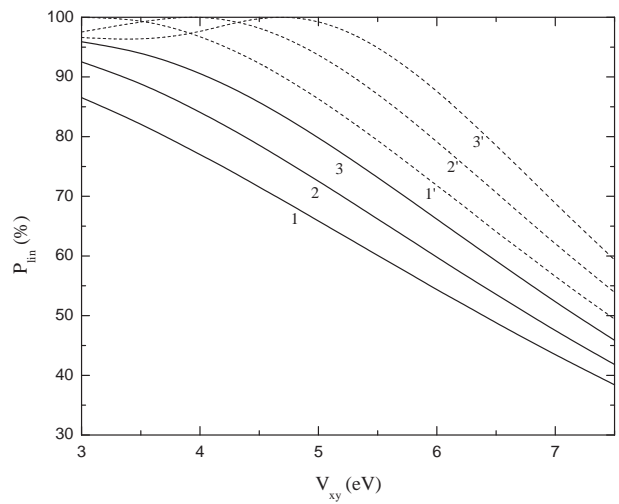}
\caption{The linear polarization of spatially-indirect optical
transitions at a ZnSe/BeTe heterointerface calculated as a
function of the interface off-diagonal tight-binding parameter
$V_{xy}$. The results are presented for the radiative
recombination between an electron from the lowest conduction
subband $e1$ and a hole from the lowest hole subband ({\it solid})
and the first excited hole subband ({\it dashed}). The calculation
is performed for three values of interface-atoms offset, $V' =
0.5$~eV (curves $1$ and $1'$), $0.75$~eV ($2$ and $2'$) and $1$~eV
($3$ and $3'$).
  }\label{f_BT_Pol_Vxy}
\end{figure}
Now we discuss the intensity $I_z$ of the $z$-polarized light
emitted in the direction $x$ or $y$ perpendicular to $z$. First of
all, we note that the optical transitions involving the heavy-hole
states become allowed in the polarization ${\bm e} \parallel z$
only due to an admixture of the $Z$ orbitals to these states.
Neglecting the spin-orbit interaction, the orbital $Z$ and the
orbitals $X,Y$ do not mix in the states with $k_x = k_y =
0$.\cite{JETP} For large values of the spin-orbit splitting of the
valence band as in the case of AlSb, an admixture of the $Z$
orbitals in the heavy-hole subband appears (i) due to the
heavy-light-hole mixing at the interfaces\cite{kamins,toropov}
or/and (ii) due to $k_{x,y}$-induced mixing of the heavy- and
light-hole states. Our calculation shows that, for an ideal
InAs/AlSb interface, the indirect radiative recombination of an
electron and a heavy hole with $k_x=k_y=0$ in the polarization
${\bm e} \parallel z$ is at least by one order of magnitude less
intensive that for ${\bm e} \perp z$. In Ref.~\onlinecite{berlin},
for the light emission along $x$ and $y$, the photoluminescence
intensity $I_z$ was small compared to $I_y$ and exceeded $I_x$ by
a factor of $\sim 2$. The observed remarkable $z$-polarized
intensity may be related to the structure imperfections and the
light depolarization on leaving the sample in the in-plane
directions.
\subsection{Lateral optical anisotropy of ZnSe/BeTe heterostructures}
In Ref.~\onlinecite{JETP} we have demonstrated that the simplest
$sp^3$ nearest-neighbor tight-binding model is consistent with the
giant linear polarization $P_{\rm lin} = 70$-$80$\% of the
vertical photoluminescence observed in (001)-grown ZnSe/BeTe
multilayered heterostructures. Here we have extended the model
from $sp^3$ to $sp^3s^*$ and included the spin-orbit interaction
into consideration. The parameters used in the calculation are
presented in Sec.~II.C except for the interface transfer integral
$V_{xy}$ and the interface-atoms offset $V'$ which were considered
as variable parameters. In the studied range of these parameters
there is no hole interface localization, and even the lowest hole
states are quantum-confined within the whole BeTe layer.

Figure 6 depicts the polarization related to the radiative
recombination at one particular interface in a 60\AA/60\AA\mbox{}
ZnSe/BeTe structure with the ZnTe-like interfaces. The
polarization degree $P_{\rm lin}$ is shown as function of $V_{xy}$
for three different values of $V'$ and for the optical transitions
involving the lowest hole subbands. As compared with the InAs/AlSb
heteropair (Figs.~\ref{f_IS_Pol_offset}, \ref{f_IS_Pol_Vxy}), the
values of $P_{\rm lin}$ are higher, and the sensitivity of $P_{\rm
lin}$ to the variation of $V_{xy}$ is much weaker. Clearly, there
exist a wide two-parametrical area of $V_{xy}$ and $V'$ where the
polarization exceeds 80\%. As well as in Ref.~\onlinecite{JETP}
the polarization sign follows the Zn-Te interface bond direction.

For comparison, we have also performed the calculations for
heterostructures with the BeSe interface atoms. The main result is
that, within the studied values of the interface parameters
$V_{xy}$ and $V'$, the polarization $P_{\rm lin}$ can reverse its
sign and is not determined completely by the interface bond
direction.
\section{Conclusion}
A tight-binding approach has been developed in order to calculate
the electronic and optical properties of type-II heterostructures.
In agreement with the existing experiments, the theory allows a
giant in-plane linear polarization for the photoluminescence of
type-II (001)-grown multi-layered structures with no-common
cations and anions, such as InAs/AlSb and ZnSe/BeTe. The
calculation shows that the high polarization can be found for the
radiative recombination involving both the interface and
quantum-confined hole states in the InAs/AlSb structures. Among
the set of interface tight-binding parameters the most important
are the interface-atoms offset $V'$ and the transfer integral
$V_{xy}$, the first controlling the hole localization at an
interface and the second controlling the in-plane optical
anisotropy. The developed theory can be generalized to calculate
the optical anisotropy of type-II quantum-wire and quantum-dot
nanostructures \cite{Toropov_ssc} to study the composition and
quality of the interfaces therein.

\acknowledgments{ This work is financially supported by the RFBR
and the programmes of RAS. One of us (M.O.N.) is grateful to the
``Dynasty'' Foundation --- ICFPM.}

\bibliography{bibliography}
\end{document}